\documentclass[
reprint,
dcolumn,
bm,
nofootinbib,
notitlepage]{revtex4-2}
\usepackage{lmodern}
\usepackage{graphicx}
\usepackage{color}
\usepackage[latin1]{inputenc}
\usepackage[T1]{fontenc}
\usepackage{amsmath}
\usepackage{amssymb}
\usepackage{braket}
\usepackage{mathrsfs}
\usepackage{hyperref}
\usepackage{ulem}
\usepackage{upgreek}
\usepackage{pstricks-add}
\usepackage{natbib} 
\usepackage{listings}
\usepackage{tikz}
\usepackage{tcolorbox}
\usetikzlibrary{mindmap}
\usepackage{pifont}
\definecolor{lightblue}{rgb}{0.145,0.6666,1}

\begin{document}
\title{The coherent-state transformation in quantum electrodynamics coupled cluster theory}

\author{Eric W. Fischer}
\email{ericwfischer.sci@posteo.de}
\affiliation{Humboldt-Universit\"at zu Berlin, Institut f\"ur Chemie, Brook-Taylor-Stra\ss e 2, D-12489 Berlin, Germany}

\date{\today}

\let\newpage\relax

\begin{abstract}
We analyse the coherent-state (CS) transformation in quantum electrodynamics coupled cluster (QED-CC) theory from the perspective of its non-vanishing commutator with the polaritonic cluster operator. Specifically, we show that a QED Hartree-Fock (QED-HF) reference state parametrized by the CS transformation leads to a QED-CC Lagrangian formally determined by CS-representations of polaritonic Hamiltonian, polaritonic cluster \textit{and} polaritonic deexcitation operators. Moreover, the herein proposed approach differs from the original formulation of QED-CC theory in the definition of the photon state basis and exploits photon-added coherent states in contrast to previously considered displaced number states. We find a renormalization of both QED-CC correlation energy and QED-CC ground state induced by the CS transformation, which depends on the mean-field expectation value of the molecular dipole operator and therefore breaks origin invariance for charged systems. Electronic contributions to correlation energy and QED-CC ground state are renormalized by CS-transformed mixed excitation and deexcitation operators. In contrast, the CS-transformed single-photon excitation affects only the QED-CC ground state but not directly the correlation energy. The renormalized QED-CC ansatz becomes similar to the original QED-CC formulation for large cavity frequencies leading to small renormalization corrections. A divergent correlation energy for molecules with a non-vanishing molecular dipole moment is found in the low-frequency limit, which we discuss with respect to multi-photon excitations in the polaritonic cluster operator and the relevance of the cavity-Born-Oppenheimer framework.
\end{abstract}

\let\newpage\relax
\maketitle
\newpage

\section{Introduction}
The experimental observation of strong light-matter coupling between confined field modes of a Fabry-P\'erot cavity and electronic excitations in organic molecules\cite{schwartz2011,hutchison2012} sparked the emerging field of polaritonic chemistry\cite{ebbesen2016,garciavidal2021}. From a theoretical perspective, the so realized electronic strong coupling (ESC) regime resembles a mixed electron-photon many-body problem, which lies beyond the capabilities of \textit{ab initio} quantum chemistry ``merely'' concerned with electronic degrees of freedom. Early efforts addressed this conceptual challenge via an extension of density functional theory (DFT) to the strong coupling regime named quantum electrodynamics DFT (QEDFT)\cite{ruggenthaler2011,tokatly2013,ruggenthaler2014,flick2015,ruggenthaler2017}, followed by more recent work on augmented Hartree-Fock (HF) and coupled cluster (CC) theories tackling the mixed fermion-boson problem from a wave function perspective\cite{haugland2020,mordovina2020}. This development consequently led to the rapidly evolving field of \textit{ab initio} polaritonic chemistry.\cite{haugland2021,deprince2021,mctague2022,riso2022,bauer2023,schnappinger2023,angelico2023,cui2024,vu2024,fischer2024,monzel2024,matousek2024,elmoutaoukal2025,fischer2025,alessandro2025} 

From the perspective of wave function theory, especially quantum electrodynamics HF (QED-HF) and CC (QED-CC) methods presented by Koch and coworkers were particularly successful.\cite{haugland2020} A central element of those approaches is an impactful mean-field product ansatz composed of a HF determinant and a coherent-state (CS) transformed photon vacuum state, which renders the mean-field energy optimal in the cavity subspace and manifestly origin-invariant.\cite{haugland2020} The QED-CC ansatz exploits the CS-transformed polaritonic (Pauli-Fierz) Hamiltonian\cite{haugland2020}, which was recently further analysed from the perspective of orbital relaxation effects\cite{liebenthal2023} and perturbation theory\cite{roden2024}. Here we take the opportunity to address the CS transformation's role in QED-CC theory beyond the mean-field orbital picture, which has apparently not been discussed in greater detail so far. Specifically, we revisit the basics of QED-CC theory motivated by the observation that the CS transformation does not commute with the polaritonic cluster operator and present an alternative QED-CC approach exploiting a distinct set of photon states.

\section{QED-CC Basics}
The QED-CC energy as introduced by Koch and coworkers is given by\cite{haugland2020}
\begin{align}
E_\mathrm{cc}
&=
\braket{
\Phi_0,0_c
\vert
e^{-\hat{Q}}
\hat{U}^\dagger_z
\hat{H}
\hat{U}_z
e^{\hat{Q}}
\vert
\Phi_0,0_c}
\quad,
\label{eq.qedcc_energy}
\end{align}
and approximates the adiabatic ground state energy of the polaritonic Hamiltonian in dipole approximation and length gauge representation 
\begin{align}
\hat{H}
=
\hat{H}_e
+
\hbar
\omega_c
\hat{b}^\dagger_\lambda
\hat{b}_\lambda
-
g_0
\sqrt{\frac{\hbar\omega_c}{2}}
\hat{d}_\lambda
(
\hat{b}^\dagger_\lambda
+
\hat{b}_\lambda
)
+
\dfrac{g^2_0}{2}
\hat{d}^2_\lambda
\,.
\label{eq.polaritonic_hamilton}
\end{align}
The first term is the electronic Hamiltonian, $\hat{H}_e$, and the second one corresponds to a (zero-point energy-shifted) cavity Hamiltonian for a single cavity mode with polarization, $\lambda$, and frequency, $\omega_c$, given in terms of photonic creation and annihilation operators, $\hat{b}^\dagger_\lambda$ and $\hat{b}_\lambda$, respectively. The remaining two contributions correspond to light-matter interaction and dipole self-energy (DSE) terms with interaction strength, $g_0$, and polarization-projected molecular dipole operator, $\hat{d}_\lambda=\hat{d}^{(e)}_\lambda+\hat{d}^{(n)}_\lambda$, composed of electronic and nuclear components, $\hat{d}^{(e)}_\lambda$ and $\hat{d}^{(n)}_\lambda$, respectively. The QED-CC wave function ansatz for Eq.\eqref{eq.qedcc_energy} reads\cite{haugland2020}
\begin{align}
\ket{\Psi^\mathrm{qed}_\mathrm{cc}}
=
e^{\hat{Q}}
\ket{\Phi_0,0_c}
\quad,
\label{eq.qedcc_ansatz}
\end{align}
with mean-field reference state, $\ket{\Phi_0,0_c}$, for the \textit{CS-transformed} polaritonic Hamiltonian, $\hat{U}^\dagger_z\hat{H}\hat{U}_z$. The mean-field reference resembles a product state between the HF-determinant, $\ket{\Phi_0}$, and the cavity mode vacuum, $\ket{0_c}$. The polaritonic cluster operator is given by\cite{haugland2020,monzel2024}
\begin{align}
\hat{Q}
&=
\hat{T}
+
\hat{\Gamma}
+
\hat{S}
\quad,
\label{eq.polaritonic_cluster_operator}
\end{align}
with electronic, $\hat{T}$, photonic, $\hat{\Gamma}$, and mixed contributions, $\hat{S}$, respectively, and allows to account for both electron and electron-photon correlation in the polaritonic ground state problem not captured on a mean-field level of theory.

\section{The CS Transformation in QED-CC}
The CS transformation enters Eq.\eqref{eq.qedcc_energy} via the bosonic operator\cite{haugland2020}
\begin{align}
\hat{U}_z
=
e^{z_\lambda(\hat{b}^\dagger_\lambda-\hat{b}_\lambda)}
\quad,
\quad
z_\lambda
=
\dfrac{
g_0
\braket{
\hat{d}_\lambda
}_0
}{\sqrt{2\hbar\omega_c}}
\quad,
\label{eq.cs_transform}
\end{align}
which depends on the mean-field dipole expectation value, $\braket{\hat{d}_\lambda}_0=\braket{\Phi_0\vert\hat{d}_\lambda\vert\Phi_0}$, and exclusively acts on the cavity subspace (\textit{cf.} Appendix \ref{sec.details_cstrafo} for details). The QED-HF reference state for the polaritonic Hamiltonian in Eq.\eqref{eq.polaritonic_hamilton} is then defined as
\begin{align}
\ket{R}
=
\hat{U}_z
\ket{\Phi_0,0_c}
\quad,
\label{eq.qedhf_ansatz}
\end{align}
with corresponding QED-HF energy 
\begin{align}
E_0
=
\braket{
\Phi_0,
0_c
\vert
\hat{U}^\dagger_z
\hat{H}
\hat{U}_z
\vert
\Phi_0,
0_c}
\quad.
\label{eq.qedhf_energy}
\end{align}
The CS transformation minimizes $E_0$ with respect to variations of photonic degrees of freedom and renders the mean-field energy origin-invariant.\cite{haugland2020} 

By inspecting Eq.\eqref{eq.qedhf_energy}, we may alternatively follow the argument of Ref.\cite{haugland2020} that $\ket{\Phi_0,0_c}$ approximates the ground state of the CS-transformed polaritonic Hamiltonian, $\hat{U}^\dagger_z\hat{H}\hat{U}_z$. Both perspectives are obviously equivalent for the mean-field energy but differ already for other mean-field expectation values, which require the CS transformation to parametrize the QED-HF reference state for properly transformed operators. We shall now extend this argument to the QED-CC approach and formulate an alternative polaritonic ground state approximation via the mean-field reference in Eq.\eqref{eq.qedhf_ansatz} given by
\begin{align}
\ket{\tilde{\Psi}^\mathrm{qed}_\mathrm{cc}}
=
&
e^{\hat{Q}}
\hat{U}_z
\ket{\Phi_0,0_c}
\neq
\hat{U}_z
e^{\hat{Q}}
\ket{\Phi_0,0_c}
=
\hat{U}_z
\ket{\Psi^\mathrm{qed}_\mathrm{cc}}
.
\label{eq.cs_qedcc_ansatz}
\end{align}
On the left-hand side, the exponentiated polaritonic cluster operator acts on the mean-field state in CS-representation in contrast to the right-hand side, where the order of operators is interchanged (\textit{cf.} Eq.\eqref{eq.qedcc_ansatz}). We like to stress that the CS-transformation is explicitly included on the right-hand side since the inequality between both expressions results from the non-vanishing commutator
\begin{align}
[e^{\hat{Q}},\hat{U}_z]
\neq
0
\quad,
\label{eq.commutator}
\end{align}
between polaritonic cluster and CS transformation operators. The energy expression for the alternative QED-CC ansatz, $\ket{\tilde{\Psi}^\mathrm{qed}_\mathrm{cc}}$, in Eq.\eqref{eq.cs_qedcc_ansatz} is given by
\begin{align}
\tilde{E}_\mathrm{cc}
&=
\braket{
\Phi_0,0_c
\vert
\hat{U}^\dagger_z
e^{-\hat{Q}}
\hat{H}
e^{\hat{Q}}
\hat{U}_z
\vert
\Phi_0,0_c}
\quad,
\label{eq.qedcc_energy_renorm}
\end{align}
and accounts for the CS transformation of \textit{both} polaritonic cluster operator \textit{and} polaritonic Hamiltonian. Thus \textit{all} photonic operators contributing to $\hat{H}$ and $\hat{Q}$ in Eq.\eqref{eq.qedcc_energy_renorm} are consistently transformed to the CS-picture in contrast to Eq.\eqref{eq.qedcc_energy}, where only photonic operators in the polaritonic Hamiltonian were subject to the CS transformation. Moreover, since the CS transformation enters the theory via the mean-field ansatz Eq.\eqref{eq.qedhf_ansatz}, the left-hand side of Eq.\eqref{eq.cs_qedcc_ansatz} seems to follow naturally for the QED-CC approach while the right-hand side can be understood as an approximation (\textit{cf.} Appendix \ref{sec.renorm_qedcc_ansatz}). We shall later discuss an alternative perspective on the original QED-CC ansatz (\textit{cf.} Sec.\ref{sec.photon_basis_states}), which provides additional insight into the difference between both approaches.

\section{Renormalized QED-CC}
In the remainder of this work, we will discuss the relevance of the non-vanishing commutator in Eq.\eqref{eq.commutator} for a QED-CC approach formulated with respect to $\ket{\tilde{\Psi}^\mathrm{qed}_\mathrm{cc}}$. To this end, we introduce a QED-CC Lagrangian 
\begin{align}
\mathcal{\tilde{L}}_\mathrm{cc}
&=
\tilde{E}_\mathrm{cc}
+
\lambda_\nu
\tilde{R}^\nu_\mathrm{cc}
\quad,
\label{eq.qed_cc_lagrangian_renorm}
\end{align}
where $\tilde{E}_\mathrm{cc}$ was given by Eq.\eqref{eq.qedcc_energy_renorm} and the second term reads 
\begin{align}
\lambda_\nu
\tilde{R}^\nu_\mathrm{cc}
&=
\braket{
\Phi_0,0_c
\vert
\hat{U}^\dagger_z
\hat{\Lambda}_Q
e^{-\hat{Q}}
\hat{H}
e^{\hat{Q}}
\hat{U}_z
\vert
\Phi_0,0_c}
\quad,
\label{eq.qedcc_residual}
\end{align}
with polaritonic deexcitation operator
\begin{align}
\hat{\Lambda}_Q
&=
\hat{T}^\dagger
+
\hat{\Gamma}^\dagger
+
\hat{S}^\dagger
\quad.
\label{eq.polaritonic_deexcitation_operator}
\end{align}
QED-CC amplitude equations, $\tilde{R}^\nu_\mathrm{cc}=0$, are obtained via the condition, $\partial_{\lambda_\nu}\mathcal{\tilde{L}}_\mathrm{cc}=0$, with multipliers, $\lambda_\nu$, for electronic, photonic and mixed excitations. We assume here and throughout the rest of this work implicit summation over repeated indices.

As an illustrative example, we discuss the QED-CCS-1-S1 scheme\cite{haugland2020,monzel2024}, which is the simplest QED-CC approximation fully accounting for electron-photon correlation in the energy expression. The corresponding polaritonic cluster operator reads
\begin{align}
\hat{Q}
=
\hat{T}_1
+
\hat{\Gamma}_1
+
\hat{S}^1_1
=
t^a_i
\hat{E}_{ai}
+
\gamma_\lambda
\hat{b}^\dagger_\lambda
+
s^\lambda_{ai}
\hat{E}_{ai}
\hat{b}^\dagger_\lambda
\quad,
\label{eq.qed_ccs_1_s1_cluster_op}
\end{align}
with single excitations in both electronic ($\hat{T}_1$) and photonic ($\hat{\Gamma}_1$) subspaces besides a mixed doubles contribution, ($\hat{S}^1_1$). Related amplitudes read $t^a_i,\gamma_\lambda$ and $s^\lambda_{ai}$, while electronic excitation operators, $\hat{E}_{ai}$, connect occupied and virtual molecular orbital (MO) subspaces characterized by indices $i,j$ and $a,b$, respectively. The polaritonic deexcitation operator follows from Eq.\eqref{eq.polaritonic_deexcitation_operator} as
\begin{align}
\hat{\Lambda}_Q
&=
\lambda^i_a
\hat{E}_{ia}
+
\gamma^\lambda
\hat{b}_\lambda
+
s^{ia}_\lambda
\hat{E}_{ia}
\hat{b}_\lambda
\quad,
\label{eq.qed_ccs_1_s1_deexc_op}
\end{align}
with electronic, photonic and mixed multipliers, $\lambda^a_i,\gamma^\lambda$ and $s^{ia}_\lambda$, while $\hat{E}_{ai}$ corresponds to an electronic \textit{deexcitation} operator. An extension to the more general QED-CCSD-1-SD1 approach as introduced in Ref.\cite{haugland2020} is provided in Appendix \ref{sec.diagrams}.

\subsection{Energy Renormalization}
We rewrite the QED-CC energy in Eq.\eqref{eq.qedcc_energy_renorm} now as
\begin{align}
\tilde{E}_\mathrm{cc}
&=
\braket{
\Phi_0,0_c
\vert
e^{-\hat{Q}_z}
\hat{H}_z
e^{\hat{Q}_z}
\vert
\Phi_0,0_c}
\quad,
\label{eq.cs_qedcc_energy_renorm}
\end{align}
where the CS-transformed polaritonic Hamiltonian, $\hat{H}_z=\hat{U}^\dagger_z\hat{H}\hat{U}_z$, is explicitly given by
\begin{multline}
\hat{H}_z
=
\hat{H}_e
+
\hbar
\omega_c
\hat{b}^\dagger_\lambda
\hat{b}_\lambda
\\
-
g_0
\sqrt{\frac{\hbar\omega_c}{2}}
\Delta\hat{d}^{(e)}_\lambda
\left(
\hat{b}^\dagger_\lambda
+
\hat{b}_\lambda
\right)
+
\dfrac{g^2_0}{2}
(\Delta\hat{d}^{(e)}_\lambda)^2
\quad,
\label{eq.cs_polaritonic_hamilton}
\end{multline}
with
\begin{align}
\Delta\hat{d}^{(e)}_\lambda
=
\hat{d}^{(e)}_\lambda
-
\braket{
\hat{d}^{(e)}_\lambda
}_0
\quad.
\label{eq.edip_corr}
\end{align}
The CS-transformed polaritonic cluster operator reads
\begin{align}
\hat{Q}_z
&=
\hat{U}^\dagger_z
\hat{Q}
\hat{U}_z
=
\hat{T}_1
+
\hat{U}^\dagger_z
\hat{\Gamma}_1
\hat{U}_z
+
\hat{U}^\dagger_z
\hat{S}^1_1
\hat{U}_z
\quad,
\label{eq.cs_polaritonic_cluster_op}
\end{align}
where the electronic single-excitation operator, $\hat{T}_1$, is invariant with respect to the CS transformation while photonic and mixed excitation operators transform as
\begin{align}
\hat{U}^\dagger_z
\hat{\Gamma}_1
\hat{U}_z
&=
\tilde{\gamma}_\lambda
\left(
\hat{b}^\dagger_\lambda
+
\dfrac{
g_0
\braket{
\hat{d}_\lambda
}_0
}{
\sqrt{2\hbar\omega_c}
}
\right)
\quad,
\label{eq.cs_photon_amplitude}
\vspace{0.2cm}
\\
\hat{U}^\dagger_z
\hat{S}^1_1
\hat{U}_z
&=
\tilde{s}^\lambda_{ai}
\hat{E}_{ai}
\left(
\hat{b}^\dagger_\lambda
+
\dfrac{
g_0
\braket{
\hat{d}_\lambda
}_0
}{
\sqrt{2\hbar\omega_c}
}
\right)
\quad.
\label{eq.cs_s11_amplitude}
\end{align}
We distinguish here photonic and mixed doubles amplitudes in the CS-representation, $\tilde{\gamma}_\lambda$ and $\tilde{s}^\lambda_{ai}$, from their original counterparts in Eq.\eqref{eq.qed_ccs_1_s1_cluster_op} due to respectively corrected amplitude equations to be discussed in Sec.\ref{sec.state_renorm}. The CS-induced shift of $\hat{S}^1_1$ can be interpreted as \textit{renormalization} of electronic singles amplitudes
\begin{align}
\tilde{t}_{ai}
&=
\tilde{t}^a_i
+
\dfrac{
g_0
\braket{
\hat{d}_\lambda
}_0
}{
\sqrt{2\hbar\omega_c}
}
\tilde{s}^\lambda_{ai}
=
\tilde{t}^a_i
+
\tilde{s}^a_i
\quad,
\label{eq.renorm_singles}
\end{align}
which depend explicitly on the light-matter interaction strength, $g_0$, cavity frequency, $\omega_c$, and molecular mean-field dipole expectation value, $\braket{\hat{d}_\lambda}_0$. Bare electronic singles amplitudes are written in the renormalized scenario as $\tilde{t}^a_i$ in line with $\tilde{\gamma}_\lambda$ and $\tilde{s}^\lambda_{ai}$. In Eq.\eqref{eq.renorm_singles}, mixed doubles amplitudes, $\tilde{s}^\lambda_{ai}$, are contracted with $\braket{\hat{d}_\lambda}_0$ via the polarization index, $\lambda$, leading to an effective electronic singles amplitude, $\tilde{s}^a_i$ (\textit{cf.} Appendix \ref{sec.diagrams} for diagrammatic representation). 

\begin{figure}[hbt!]
\includegraphics[scale=1.0]{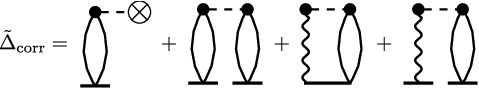}
\caption{Diagrams representing the QED-CCS-1-S1 correlation correction in Eq.\eqref{eq.qed_cc_corr}.}
\label{fig.qedccs_1_s1_energy}
\end{figure}

The renormalized QED-CC correlation energy is now obtained as
\begin{align}
\Delta
\tilde{E}_\mathrm{corr}
&=
\tilde{\Delta}_\mathrm{corr}
+
\tilde{\Delta}_\mathrm{ren}
\quad,
\label{eq.cs_qed_cc_corr}
\end{align} 
where the QED-CCS-1-S1 correlation correction reads explicitly
\begin{multline}
\tilde{\Delta}_\mathrm{corr}
=
\tilde{t}^a_i
\tilde{f}_{ai}
+
\dfrac{1}{2}
\tilde{t}^a_i
\tilde{t}^b_j
\bar{w}^{ai}_{bj}
\\
-
g_0
\sqrt{\dfrac{\hbar\omega_c}{2}}
\left(
\tilde{s}^\lambda_{ai}
+
\tilde{t}^a_i
\tilde{\gamma}_\lambda
\right)
d^{ai}_\lambda
\quad,
\label{eq.qed_cc_corr}
\end{multline}
with $\tilde{f}_{ai}=\braket{\Phi^a_i,0_c\vert\hat{H}_z\vert\Phi_0,0_c}$ and DSE-augmented antisymmetrized two-electron integrals
\begin{align}
\bar{w}^{ai}_{bj}
&=
\tilde{g}_{aibj}
-
\tilde{g}_{ajbi}
\quad,
\end{align} 
where $\tilde{g}_{aibj}=g_{aibj}+g^2_0d^{ai}_\lambda d^{bj}_\lambda$ contains bare two-electron integrals in Mulliken notation, $g_{aibj}=(ai\vert bj)$, and polarization-projected transition dipole matrix elements, $d^{ai}_\lambda$ and $d^{bj}_\lambda$, respectively. The first term of Eq.\eqref{eq.qed_cc_corr} vanishes for canonical QED-HF MOs satisfying Brillouin's theorem, $\tilde{f}_{ai}=0$. In Fig.\ref{fig.qedccs_1_s1_energy}, we provide the diagrammatic representation of Eq.\eqref{eq.qed_cc_corr} as introduced in Ref.\cite{monzel2024}.

\begin{figure}[hbt!]
\includegraphics[scale=1.0]{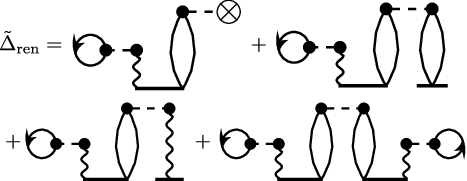}
\caption{Diagrams representing the QED-CCS-1-S1 renormalization correction in Eq.\eqref{eq.qed_cc_renorm}.}
\label{fig.qedccs_1_s1_renorm}
\end{figure}

The CS-induced renormalization correction is obtained as 
\begin{multline}
\tilde{\Delta}_\mathrm{ren}
=
\tilde{s}^a_i
\tilde{f}_{ai}
+
\dfrac{1}{2}
\left(
\tilde{s}^a_i
\tilde{t}^b_j
+
\tilde{s}^b_j
\tilde{t}^a_i
\right)
\bar{w}^{ai}_{bj}
\\
-
g_0
\sqrt{\dfrac{\hbar\omega_c}{2}}
\tilde{s}^a_i
\tilde{\gamma}_{\lambda}
d^{ai}_{\lambda}
+
\dfrac{1}{2}
\tilde{s}^a_i
\tilde{s}^b_j
\bar{w}^{ai}_{bj}
\quad,
\label{eq.qed_cc_renorm}
\end{multline}
where we exploit effective singles amplitudes, $\tilde{s}^a_i$, introduced in Eq.\eqref{eq.renorm_singles} with diagrammatric representation given in Fig.\ref{fig.qedccs_1_s1_renorm}.
The first term vanishes also here in case of canonical QED-HF MOs due to Brillouin's theorem. Moreover, we obtain two contributions from the disconnected electronic doubles correction determined by DSE-augmented two-electron integrals, $\bar{w}^{ai}_{bj}$, and additionally a third term resulting from the disconnected mixed doubles contribution to the electron-photon correlation energy. In Appendix \ref{sec.diagrams}, we provide additional corrections relevant for the QED-CCSD-1-SD1 scheme besides details on bubble contractions.

We focused so far on the CS-transformed mixed-doubles excitation operator in Eq.\eqref{eq.cs_s11_amplitude} and its connection to electronic single excitations. The single photon counterpart in Eq.\eqref{eq.cs_photon_amplitude} acquires in contrast only a constant shift, which vanishes throughout the Baker-Campbell-Hausdorff (BCH) expansion of the similarity-transformed polaritonic Hamiltonian in CS-representation and consequently does not alter the QED-CC energy. However, the CS transformation does alter the photonic single-excitation component of the QED-CC state as we shall address in the following. We shall show in Sec.\ref{sec.two_photon} how two-photon excitations\cite{pavosevic2022,philbin2023,haugland2025} effectively renormalize single photon amplitudes and therefore alter the correlation energy.

\subsection{State Renormalization}
\label{sec.state_renorm}
We turn our attention now to the QED-CC amplitude equations obtained from Eq.\eqref{eq.qedcc_residual}. Under the action of $\hat{U}_z$, we find photonic contributions of the polaritonic deexcitation operator in Eq.\eqref{eq.qed_ccs_1_s1_deexc_op} to transform as
\begin{align}
\hat{U}^\dagger_z
\hat{\Gamma}^\dagger_1
\hat{U}_z
&=
\tilde{\gamma}^\lambda
\left(
\hat{b}_\lambda
+
\dfrac{
g_0
\braket{
\hat{d}_\lambda
}_0
}{
\sqrt{2\hbar\omega_c}
}
\right)
\quad,
\label{eq.cs_photon_multiplier}
\vspace{0.2cm}
\\
\hat{U}^\dagger_z
(\hat{S}^1_1)^\dagger
\hat{U}_z
&=
\tilde{s}^{ia}_\lambda
\hat{E}_{ia}
\left(
\hat{b}_\lambda
+
\dfrac{
g_0
\braket{
\hat{d}_\lambda
}_0
}{
\sqrt{2\hbar\omega_c}
}
\right)
\quad,
\label{eq.cs_s11_multiplier}
\end{align}
leading to a renormalization of electronic singles multipliers due to the second term in Eq.\eqref{eq.cs_s11_multiplier} as 
\begin{align}
\tilde{\lambda}^{ia}
&=
\tilde{\lambda}^i_a
+
\dfrac{
g_0
\braket{
\hat{d}_\lambda
}_0
}{
\sqrt{2\hbar\omega_c}
}
\tilde{s}^{ia}_\lambda
\label{eq.renorm_singles_amp}
\quad,
\end{align}
in analogy to the related electronic singles amplitudes in Eq.\eqref{eq.renorm_singles}. We may now write Eq.\eqref{eq.qedcc_residual} for the QED-CCS-1-S1 scheme as 
\begin{align}
\lambda_\nu
\tilde{R}^\nu_\mathrm{cc}
&=
\tilde{\lambda}^i_a
\tilde{R}^a_i
+
\tilde{\gamma}^\lambda
\tilde{R}_\lambda
+
\tilde{s}^{ia}_\lambda
\tilde{R}^\lambda_{ai}
\quad,
\end{align}
and obtain related amplitude equations by demanding $\mathcal{\tilde{L}}_\mathrm{cc}$ to be stationary with respect to variation of multipliers $\tilde{\lambda}^i_a,\tilde{s}^{ia}_\lambda$ and $\tilde{\gamma}^\lambda$  
\begin{align}
\tilde{R}^a_i
&=
\braket{
\Phi^i_a,0_c
\vert
e^{-\hat{Q}_z}
\hat{H}_z
e^{\hat{Q}_z}
\vert
\Phi_0,0_c}
=
0
\quad,
\label{eq.singles_amplitude_renorm}
\vspace{0.2cm}
\\
\tilde{R}_\lambda
&=
\braket{
\Phi_0,1_c
\vert
e^{-\hat{Q}_z}
\hat{H}_z
e^{\hat{Q}_z}
\vert
\Phi_0,0_c}
\label{eq.photon_amplitude_renorm}
\vspace{0.2cm}
\\
&
\hspace{0.2cm}
+
\frac{
g_0
\braket{
\hat{d}_\lambda
}_0
}{
\sqrt{2\hbar\omega_c}
}
\braket{
\Phi_0,0_c
\vert
e^{-\hat{Q}_z}
\hat{H}_z
e^{\hat{Q}_z}
\vert
\Phi_0,0_c}
=
0
\nonumber
\,,
\vspace{0.2cm}
\\
\tilde{R}^\lambda_{ai}
&=
\braket{
\Phi^i_a,1_c
\vert
e^{-\hat{Q}_z}
\hat{H}_z
e^{\hat{Q}_z}
\vert
\Phi_0,0_c}
\label{eq.mixed_amplitude_renorm}
\vspace{0.2cm}
\\
&
\hspace{0.2cm}
+
\frac{
g_0
\braket{
\hat{d}_\lambda
}_0
}{
\sqrt{2\hbar\omega_c}
}
\braket{
\Phi^i_a,0_c
\vert
e^{-\hat{Q}_z}
\hat{H}_z
e^{\hat{Q}_z}
\vert
\Phi_0,0_c}
=
0
\nonumber
\,.
\end{align}
Here, we introduced singly-excited determinants, $\ket{\Phi^a_i}$, and single-photon states, $\ket{1_c}$, respectively. In contrast to the energy expression, we find a new contribution for both photon \textit{and} mixed-doubles amplitude equations resulting from the CS transformation. Thus, CS-transformed single-photon excitations directly modify the QED-CC polaritonic ground state approximation but affect the QED-CC correlation energy only indirectly via their influence on other amplitudes. As noted above, we shall see in the following discussion that this statement holds only for single-photon truncations of the polaritonic cluster operator.

\section{Discussion}
We turn now to a more detailed discussion of renormalized QED-CC and its relation to the original formalism. In a first step, we address additional properties of renormalized QED-CC energy and state expressions related to the CS-transformation parameter introduced in Eq.\eqref{eq.cs_transform}. We subsequently show how two-photon excitations and related mixed-triples components in the polaritonic cluster operator effectively renormalize single-photon, mixed-doubles and electronic singles amplitudes. We eventually discuss an alternative perspective on the original QED-CC ansatz based on a redefinition of the exponentiated cluster operator, which highlights the difference between the two approaches due to distinct definitions of photon states.

\subsection{The Renormalization Correction}
The renormalization correction, $\tilde{\Delta}_\mathrm{ren}$, in Eq.\eqref{eq.qed_cc_renorm} depends on the molecular mean-field dipole expectation value, $\braket{\hat{d}_\lambda}_0$. Accordingly, it is only relevant for molecules with a non-vanishing permanent dipole moment, $\braket{\hat{d}_\lambda}_0\neq 0$, along the cavity-mode polarization axis, $\lambda$, on a QED-HF level of theory. Moreover, $\tilde{\Delta}_\mathrm{ren}$ breaks origin-invariance for charged molecules even in absence of a permanent dipole moment where the cavity field couples to the total molecular charge. This observation is in agreement with properties of the converged QED-HF reference state, $\ket{R}$, as already pointed out in Ref.\cite{haugland2020}. Notably, the renormalization correction can be expected to be small for \textit{large} cavity frequencies as relevant in the ESC regime such that renormalized and original QED-CC approaches become similar. We will discuss differences emerging in the \textit{low}-frequency regime in detail below (\textit{cf.} Sec.\ref{eq.low_freq}).

Turning to the polaritonic cluster operator, we have seen that the mixed-doubles contribution, $\hat{S}^1_1$, acquires a purely electronic single-excitation correction under the CS transformation, which can be interpreted as renormalization of the corresponding electronic single excitation operator, $\hat{T}_1$. Thus, the resulting QED-CC correlation energy in Eq.\eqref{eq.cs_qed_cc_corr} can be made formally equivalent to the original QED-CC expression by replacing $\tilde{t}^a_i$ in Eq.\eqref{eq.qed_cc_corr} with $\tilde{t}_{ai}$. The same holds for electronic and mixed amplitude equations, when one replaces $\tilde{\lambda}^i_a$ with the renormalized $\tilde{\lambda}^{ia}$ expression in Eq.\eqref{eq.renorm_singles_amp}. However, this is not possible for single-photon excitations due to the non-uniform behaviour of the constant shift in Eqs.\eqref{eq.cs_photon_amplitude} and \eqref{eq.cs_photon_multiplier}, resulting only in a modified amplitude equation \eqref{eq.photon_amplitude_renorm}. A consistent reformulation of the renormalized QED-CC ansatz in terms of quantities determining the original formulation seems therefore not possible.

A second aspect related to the polaritonic cluster operator concerns an apparent connection between QED-CC truncation schemes and the CS transformation. In order to illustrate this aspect, we consider the QED-CCD-1-S1 scheme with polaritonic cluster operator
\begin{align}
\hat{Q}
&=
\dfrac{1}{4}
t^{ab}_{ij}
\hat{E}^a_i
\hat{E}^b_j
+
\gamma_{\lambda}
\hat{b}^\dagger_\lambda
+
s^{ai}_\lambda
\hat{E}^a_i
\hat{b}^\dagger_\lambda
\quad,
\label{eq.ccd_1_s1}
\end{align}
where we now account for electronic \textit{double} excitations with amplitudes, $t^{ab}_{ij}$. Under the action of the CS transformation, the mixed doubles excitation operator (third term) acquires now a seemingly ``spurious'' electronic single-excitation contribution (\textit{cf.} Eq.\eqref{eq.cs_s11_amplitude}), which renders the manifold of singly-excited determinants to be relevant although not considered in the intended QED-CC approximation. Note, the similarity transformation in CC theory commonly leads to the effective inclusion of \textit{higher} excitations, while we find here lower-lying states to become relevant. This observation might suggest that a ``consistent'' truncation of the polaritonic cluster operator considers only electronic and mixed excitation operators in agreement with CS-induced renormalization corrections, \textit{e.g.}, $\hat{T}_1$ and $\hat{S}^1_1$ or $\hat{T}_2$ and $\hat{S}^1_2$ with mixed-triples excitations, $\hat{S}^1_2=\frac{1}{4}s^\lambda_{aibj}\hat{E}^a_i\hat{E}^b_j\hat{b}^\dagger_\lambda$, respectively. 

\subsection{Photon Basis States and the CS-Transformation}
\label{sec.photon_basis_states}
An alternative perspective on the original formulation of QED-CC theory emerges by exploiting unitarity of the CS-transformation as
\begin{align}
\ket{\Psi^\mathrm{qed}_\mathrm{cc}}
&\equiv
\hat{U}_z
e^{\hat{Q}}
\ket{\Phi_0,0_c}
=
e^{\hat{Q}_z}
\hat{U}_z
\ket{\Phi_0,0_c}
\quad,
\label{eq.qed_cc_alternative}
\end{align}
where we explicitly include $\hat{U}_z$ in the ansatz in contrast to Eq.\eqref{eq.qedcc_ansatz}. The right-hand side follows straightforwardly via the identity, $\hat{U}^\dagger_z\hat{U}_z=1$, in combination with definitions given in Eqs.\eqref{eq.qedhf_ansatz} and \eqref{eq.cs_polaritonic_cluster_op}, respectively. We like to stress that this result is equivalent to the right-hand side of Eq.\eqref{eq.cs_qedcc_ansatz} although both cluster operator and QED-HF reference state are now explicitly treated in CS-representation. We may now perform a BCH expansion of $e^{\hat{Q}_z}$ in Eq.\eqref{eq.qed_cc_alternative} to obtain the renormalized QED-CC ansatz as first-order approximation (\textit{cf.} Appendix \ref{sec.renorm_qedcc_ansatz} for details)
\begin{align}
\ket{\Psi^\mathrm{qed}_\mathrm{cc}}
&=
\ket{\tilde{\Psi}^\mathrm{qed}_\mathrm{cc}}
+
\left(
[\hat{G}_z,e^{\hat{Q}}]
+
\dots
\right)
\hat{U}_z
\ket{\Phi_0,0_c}
\quad.
\label{eq.irdiv_qedcc}
\end{align}
At first glance this result seems unexpected in context of our previous statement that the original QED-CC formalism can be understood as approximation of the renormalized QED-CC ansatz. We shall show now that both approaches actually resemble two distinct perspectives on the coupled electron-photon problem, which differ in their definition of photon states. In order to illustrate this aspect, we consider the QED configuration interaction (QED-CI) scheme for the sake of simplicity, where related polaritonic ground state approximations would be given by
\begin{align}
\ket{\Psi^\mathrm{qed}_\mathrm{ci}}
&=
\hat{U}_z
\left(
C_0
+
C^a_i
\hat{E}_{ai}
+
C_\lambda
\hat{b}_\lambda
+
\dots
\right)
\ket{\phi_0,0_c}
\,,
\vspace{0.2cm}
\\
\ket{\tilde{\Psi}^\mathrm{qed}_\mathrm{ci}}
&=
\left(
C_0
+
C^a_i
\hat{E}_{ai}
+
C_\lambda
\hat{b}_\lambda
+
\dots
\right)
\hat{U}_z
\ket{\phi_0,0_c}
\,,
\end{align}
with coefficients, $C_0,C^a_i$ and $C_\lambda$. We realize now that both approaches differ in the order the reference state's photon component being excited and ``displaced'': In the first line, we find the photon creation operator to act on $\ket{\phi_0,0_c}$ leading to a single-photon number state followed by displacement via the CS-transformation. The second line in contrast reflects a single-photon excitation of a coherent state. Both states are in general inequivalent since $[\hat{U}_z,\hat{b}^\dagger_\lambda]\neq 0$ and known as \textit{displaced number states}\cite{cahill1969,oliveira1990,ziesel2013} and \textit{photon-added coherent states}\cite{agarwal1991,fadrny2024}, respectively. Thus, the original QED-CC formalism exploits a displaced number state basis while the herein proposed renormalized QED-CC ansatz builds on a basis of photon-added coherent states. This observation is actually already evident from Ref.\cite{haugland2020}, however, the alternative perspective in Eq.\eqref{eq.qed_cc_alternative} allows now for a connection to the renormalized QED-CC ansatz. 

\subsection{Two-Photon Excitations and Renormalization}
\label{sec.two_photon}
\begin{figure}[hbt!]
\includegraphics[scale=1.0]{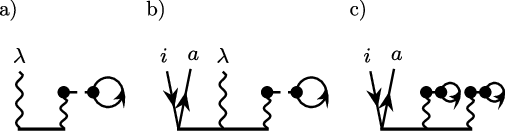}
\caption{Diagrams representing two-photon-induced renormalization corrections for a) single-photon ($\tilde{\gamma}^{(2)}_\lambda$), b) mixed-doubles ($\tilde{s}^{\lambda(2)}_{ai}$) and c) electronic single amplitudes, $\tilde{s}^{(2)}_{ai}$, as given in Eqs.\eqref{eq.eff_1p} and \eqref{eq.mixed_doubles_1e1p_singles_1e}.}
\label{fig.renorm_photon}
\end{figure}

The nature of two-photon excitations in QED-CC is significantly less studied than its single-photon counterpart but has acquired some recent attention.\cite{pavosevic2022,philbin2023,haugland2025} The corresponding excitation operator is defined by
\begin{align}
\hat{\Gamma}_2
&=
\gamma_{\lambda\lambda^\prime}
\hat{b}^\dagger_\lambda
\hat{b}^\dagger_{\lambda^\prime}
\quad,
\end{align}
with amplitude, $\gamma_{\lambda\lambda^\prime}$, and turns under the CS-transformation into
\begin{multline}
\hat{U}^\dagger_z
\hat{\Gamma}_2
\hat{U}_z
=
\tilde{\gamma}_{\lambda\lambda^\prime}
\left(
\hat{b}^\dagger_\lambda
+
\dfrac{
g_0
\braket{
\hat{d}_\lambda
}_0
}{
\sqrt{2\hbar\omega_c}
}
\right)
\left(
\hat{b}^\dagger_{\lambda^\prime}
+
\dfrac{
g_0
\braket{
\hat{d}_{\lambda^\prime}
}_0
}{
\sqrt{2\hbar\omega_c}
}
\right)
\\
=
\tilde{\gamma}_{\lambda\lambda^\prime}
\hat{b}^\dagger_\lambda
\hat{b}^\dagger_{\lambda^\prime}
+
\tilde{\gamma}^{(2)}_\lambda
\hat{b}^\dagger_\lambda
+
\dfrac{
g^2_0
\braket{
\hat{d}_{\lambda^\prime}
}^2_0
}{
2\hbar\omega_c
}
\quad.
\end{multline}
In the second term, we introduced an effective single-photon amplitude 
\begin{align}
\tilde{\gamma}^{(2)}_\lambda
=
2
\dfrac{
g_0
\braket{
\hat{d}_{\lambda^\prime}
}_0
}{
\sqrt{2\hbar\omega_c}
}
\tilde{\gamma}_{\lambda\lambda^\prime}
\quad,
\label{eq.eff_1p}
\end{align}
which constitutes a renormalization of the bare single-photon amplitude, $\tilde{\gamma}_\lambda$ (\textit{cf.} Fig.\ref{fig.renorm_photon}a for diagrammatic representation). 
Two-photon excitations may additionally contribute to mixed-triples excitations via
\begin{align}
\hat{S}^2_1
&=
\tilde{s}^{\lambda\lambda^\prime}_{ai}
\hat{E}_{ai}
\hat{b}^\dagger_\lambda
\hat{b}^\dagger_{\lambda^\prime}
\quad,
\end{align}
which transforms as
\begin{multline}
\hat{U}^\dagger_z
\hat{S}^2_1
\hat{U}_z
=
\tilde{s}^{\lambda\lambda^\prime}_{ai}
\hat{E}_{ai}
\left(
\hat{b}^\dagger_\lambda
+
\dfrac{
g_0
\braket{
\hat{d}_\lambda
}_0
}{
\sqrt{2\hbar\omega_c}
}
\right)
\left(
\hat{b}^\dagger_{\lambda^\prime}
+
\dfrac{
g_0
\braket{
\hat{d}_{\lambda^\prime}
}_0
}{
\sqrt{2\hbar\omega_c}
}
\right)
\\
=
\tilde{s}^{\lambda\lambda^\prime}_{ai}
\hat{E}_{ai}
\hat{b}^\dagger_\lambda
\hat{b}^\dagger_{\lambda^\prime}
+
\tilde{s}^{\lambda(2)}_{ai}
\hat{E}_{ai}
\hat{b}^\dagger_\lambda
+
\tilde{s}^{(2)}_{ai}
\hat{E}_{ai}
\quad,
\label{eq.mixed_triples_1e2p}
\end{multline}
with an effective mixed-doubles and a higher-order effective electronic singles amplitude (\textit{cf.} Fig.\ref{fig.renorm_photon}b and c)
\begin{align}
\tilde{s}^{\lambda(2)}_{ai}
&=
2
\dfrac{
g_0
\braket{
\hat{d}_{\lambda^\prime}
}_0
}{
\sqrt{2\hbar\omega_c}
}
\tilde{s}^{\lambda\lambda^\prime}_{ai}
\,,
\,
\tilde{s}^{(2)}_{ai}
=
\dfrac{
g^2_0
\braket{
\hat{d}_{\lambda}
}_0
\braket{
\hat{d}_{\lambda^\prime}
}_0
}{
2\hbar\omega_c
}
\tilde{s}^{\lambda\lambda^\prime}_{ai}
\,.
\label{eq.mixed_doubles_1e1p_singles_1e}
\end{align}
We note that the second-order renormalization correction or electronic singles amplitudes scales as $\tilde{s}^{(2)}_{ai}\propto\frac{1}{\omega_c}$ and is therefore less relevant than $\tilde{s}^a_i\propto\frac{1}{\sqrt{\omega_c}}$ in the high-frequency ESC regime. Consequences for the low-frequency regime will be addressed in Sec.\eqref{eq.low_freq}.

\subsection{The Low-Frequency Regime}
\label{eq.low_freq}
For large cavity frequencies, the CS-transformation parameter, $z_\lambda$, tends to be small such that the herein introduced renormalized formulation become similar to the original QED-CC approach\cite{haugland2020}. However, in the low-frequency limit, $z_\lambda\sim\frac{1}{\sqrt{\omega_c}}\to\infty$ for $\omega_c\to 0$, such that renormalization corrections diverge and the renormalized QED-CC ground state becomes undefined. This observation is challenged by recent theoretical results of Haugland \textit{et al.}\cite{haugland2025} reporting instead on a smooth convergence of the QED-CC energy in Eq.\eqref{eq.qedcc_energy} towards a frequency-independent value as $\omega_c\to 0$. The authors rationalized their findings by noting that the electron-photon correlation contribution to the QED-CC energy vanishes as $\omega_c\to 0$ in line with the related light-matter interaction term. The low-frequency limit was then interpreted as the cavity frequency-independent cavity Born-Oppenheimer (CBO) regime.\cite{haugland2025} 

We shall first note that a vanishing light-matter interaction contribution to Eq.\eqref{eq.qedcc_energy} does not necessarily lead to a stationary \textit{correlated} electronic energy in the CBO framework as discussed in Refs.\cite{angelico2023,fischer2025}. Furthermore, the CBO ground state problem is by construction ``photon-free'' and purely electronic due to adiabatic separation of electrons and cavity modes.\cite{flick2017,flick2017cbo,fischer2023}. In contrast, the (approximate) adiabatic \textit{polaritonic} ground state exhibits a divergent photon number in the low-frequency limit already on the level of QED-HF theory (\textit{cf.} Appendix \ref{sec.photonnr_qedhf})
\begin{align}
\braket{\hat{n}_\lambda}_0
&=
\dfrac{g^2_0}{2\hbar\omega_c}
\left(
\braket{
\hat{d}^{(e)}_\lambda
\hat{d}^{(e)}_\lambda
}_0
-
\braket{
\hat{d}^{(e)}_\lambda
}^2_0
\right)
-
\dfrac{1}{2}
\to
\infty
\quad,
\label{eq.cs_photon_expect}
\end{align}
as $\omega_c\to 0$, which renders a comparison with the CBO regime conceptually difficult. In addition, for molecular systems with a non-vanishing permanent dipole moment on the mean-field level of theory, $\braket{\hat{d}_\lambda}_0\neq 0$, the QED-HF reference state in Eq.\eqref{eq.qedhf_ansatz} is undefined in the low-frequency limit since all components of the coherent state vanish identically for $\omega_c\to 0$ (\textit{cf.} Appendix \ref{sec.lowfreq_qedhf}). From a formal perspective, we should expect a similar divergent behaviour of the QED-CC ansatz for consistency reasons, which is however only guaranteed when we account for the CS-transformation (\textit{cf.} Eq.\eqref{eq.cs_qedcc_ansatz}). 

We may now recall that on a \textit{mean-field} level of theory, the ground state energy is constant as $\omega_c\to 0$ despite the related reference state being undefined. Also in this case, one formally does not obtain the CBO result since mean-field reference states differ by definition and therefore other expectation values besides the energy. When we turn to the renormalized QED-CC approach, the divergent correlation energy can be understood as consequence of the undefined reference state, which carries over to photon states (\textit{cf.} Sec.\ref{sec.photon_basis_states}). For small but non-zero cavity frequencies in contrast, excited states are well defined and we know from the discussion in Sec.\ref{sec.two_photon} that multi-photon renormalization corrections become significant for the electron-photon correlation energy (\textit{cf.} Fig.\ref{fig.renorm_photon}). The relevance of two-photon excitations was also emphasized in Ref.\cite{haugland2025} despite their influence on the electron-photon correlation energy being only indirect via other amplitudes in the original QED-CC approach. Nevertheless, the relevance of higher-order photon excitations seems plausible  for small $\omega_c$ as multi-photon states energetically accumulate in the vicinity of electronic reference states. Thus, even higher-order photon states might contribute to a converged QED-CC ground state energy, potentially even beyond a two-photon truncation of the cluster operator. Such a convergence aspect concerning the truncation of the polaritonic cluster operator with respect to photon excitations seems yet to be investigated and might benefit from the QED-FCI approach, which is beyond the scope of this work. A conclusive answer concerning the convergence of the electron-photon correlation energy in the low-frequency limit may thus require to account for such multi-photon effects.

We shall finally note a more subtle problem concerning the actual definition of an adiabatic polaritonic ground state in the low-frequency limit. As discussed in Ref.\cite{fischer2023}, cavity-induced non-adiabatic interactions with the energetically low-lying excited state manifold can occur for small $\omega_c$, which challenges the interpretation of a ground state potential energy surface. In this context it might be interesting to exploit the equation-of-motion extension of QED-CC to investigate the low-frequency behaviour of excited state energies beyond the results reported by Ref.\cite{haugland2025}. Eventually, for systems with \textit{electronic} excited states being energetically well separated from the ground state, this situation is conceptually resolved via the CBO framework\cite{flick2017,flick2017cbo,fischer2023} and related implementations of CC theory\cite{angelico2023,fischer2024,fischer2025}, which therefore seem to be more suitable for addressing the low-frequency regime.

\section{Conclusion}
We analysed the coherent-state transformation in QED-CC theory from the perspective of its non-vanishing commutator with the polaritonic cluster operator. Specifically, we showed that for a CS-parametrized QED-HF reference state, the QED-CC Lagrangian is subject to CS-transformed polaritonic Hamiltonian, cluster and deexcitation operators, which we named renormalized QED-CC. The latter is subject to renormalization corrections of QED-CC energy and amplitude equations for molecules with permanent dipole moment or non-vanishing total charge, which depends on the light-matter interaction strength and the cavity frequency. Moreover, the CS transformation seemingly supports only QED-CC truncation schemes where the orders of purely electronic excitation operators and electronic components in mixed excitation operators coincide, while other schemes are subject to ``spurious'' excited state contributions. We furthermore showed that the original and the renormalized QED-CC formalism fundamentally differ in the definition of photon states, which are identified as displaced number states and photon-added coherent states. Two-photon contributions to the polaritonic cluster operator are shown to renormalize single-photon and mixed-doubles contributions while providing higher-order corrections of electronic singles amplitudes. For large cavity frequencies, renormalization corrections become small rendering both methods similar. In contrast, the low-frequency limit, $\omega_c\to 0$, is characterized by a divergence of ground state and correlation energy in renormalized QED-CC traced back to the definition of photon basis states. A recently reported smooth convergence of the QED-CC energy to a constant low-frequency limit is discussed in the context of multi-photon excitations and problems arising in the definition of a polaritonic ground state in the presence of non-adiabatic coupling to an energetically low-lying excited state manifold. The role of the cavity-Born-Oppenheimer (CBO) framework and related CBO-CC implementations is highlighted for the low-frequency regime.

Since we limited our discussion to the ground state problem, it remains to be analysed how the the CS transformation affects the equation-of-motion formulation of QED-CC theory for excited states. This aspect might be particularly interesting for addressing the low-frequency behaviour of excited states yet to be studied to the best of our knowledge. Our analysis eventually suggests that ``renormalization'' or the definition of photon states might also affect other correlation methods building upon a mean-field ansatz in CS-representation. We hope this work stimulates further discussion of basic assumptions in \textit{ab initio} polaritonic chemistry and the connection between \textit{ab initio} QED and CBO frameworks.

\section*{Acknowledgements}
E.W. Fischer acknowledges funding by the Deutsche Forschungsgemeinschaft (DFG, German Research Foundation) through DFG project 536826332 and support by Michael Roemelt.

\section*{Conflict of Interest}
The author has no conflicts to disclose.

\section*{Data Availability Statement}
Data sharing is not applicable to this article as no new data were created or analysed in this study.

\renewcommand{\thesection}{}
\section*{Appendix}

\setcounter{equation}{0}
\renewcommand{\theequation}{\thesubsection.\arabic{equation}}
\subsection{Details of the CS Transformation}
\label{sec.details_cstrafo}
We derive the CS transformation parameter, $z_\lambda$, in Eq.\eqref{eq.cs_transform} such that the respective polaritonic Hamiltonian in CS-representation (\textit{cf.} Eq.\eqref{eq.cs_polaritonic_hamilton}) is obtained. We first note that bosonic operators transform for a real $z_\lambda$ as
\begin{align}
\hat{U}^\dagger_z
\hat{b}^\dagger_\lambda
\hat{U}_z
=
\hat{b}^\dagger_\lambda
+
z_\lambda
\quad,
\quad
\hat{U}^\dagger_z
\hat{b}_\lambda
\hat{U}_z
=
\hat{b}_\lambda
+
z_\lambda
\quad.
\end{align}
The relevant terms in the polaritonic Hamiltonian (\textit{cf.} Eq.\eqref{eq.polaritonic_hamilton}), which transform non-trivially under the action of $\hat{U}_z$, are the cavity Hamiltonian and the light-matter interaction term
\begin{align}
\hat{H}_c
&=
\hbar\omega_c
\hat{b}^\dagger_\lambda
\hat{b}_\lambda
\quad,
\vspace{0.2cm}
\\
\hat{H}_\mathrm{int}
&=
-
g_0
\sqrt{\dfrac{\hbar\omega_c}{2}}
\hat{d}_\lambda
(
\hat{b}^\dagger_\lambda
+
\hat{b}_\lambda
)
\quad,
\end{align}
which turn into
\begin{align}
\hat{U}^\dagger_z
\hat{H}_c
\hat{U}_z
=
\hbar\omega_c
\hat{b}^\dagger_\lambda
\hat{b}_\lambda
+
\hbar\omega_c
z_\lambda
\left(
\hat{b}^\dagger_\lambda
+
\hat{b}_\lambda
\right)
+
\hbar\omega_c
z^2_\lambda
\,,
\end{align}
and
\begin{multline}
\hat{U}^\dagger_z
\hat{H}_\mathrm{int}
\hat{U}_z
=
-
g_0
\sqrt{\dfrac{\hbar\omega_c}{2}}
\hat{d}_\lambda
(
\hat{b}^\dagger_\lambda
+
\hat{b}_\lambda
)
\\
-
g_0
\sqrt{2\hbar\omega_c}
\hat{d}_\lambda
z_\lambda
\quad.
\end{multline}
Light-matter interaction and dipole self-energy terms in CS-representation are obtained by collecting $z_\lambda$-dependent terms as
\begin{align}
\tilde{H}_\mathrm{int}
&=
-
g_0
\sqrt{\dfrac{\hbar\omega_c}{2}}
\left(
\hat{d}_\lambda
-
\dfrac{\sqrt{2\hbar\omega_c}}{g_0}
z_\lambda
\right)
(
\hat{b}^\dagger_\lambda
+
\hat{b}_\lambda
)
\quad,
\vspace{0.2cm}
\\
\tilde{H}_\mathrm{dse}
&=
\dfrac{g^2_0}{2}
\hat{d}^2_\lambda
-
g_0
\sqrt{2\hbar\omega_c}
\hat{d}_\lambda
z_\lambda
+
\hbar\omega_c
z^2_\lambda
\quad.
\end{align}
The CS transformation parameter is determined via
\begin{align}
\hat{d}_\lambda
-
\dfrac{\sqrt{2\hbar\omega_c}}{g_0}
z_\lambda
=
\hat{d}^{(e)}_\lambda
-
\braket{\hat{d}^{(e)}_\lambda}_0
\quad,
\end{align}
which leads to the desired result
\begin{align}
z_\lambda
&=
\dfrac{g_0\braket{\hat{d}_\lambda}_0}{\sqrt{2\hbar\omega_c}}
>
0
\quad,
\end{align}
such that the DSE term follows as
\begin{align}
\tilde{H}_\mathrm{dse}
&=
\dfrac{g^2_0}{2}
\hat{d}^2_\lambda
-
g^2_0
\hat{d}_\lambda
\braket{\hat{d}_\lambda}_0
+
\dfrac{g^2_0}{2}
\braket{\hat{d}_\lambda}^2_0
\quad,
\vspace{0.2cm}
\\
&=
\dfrac{g^2_0}{2}
\left(
\hat{d}_\lambda
-
\braket{\hat{d}_\lambda}_0
\right)^2
\quad,
\vspace{0.2cm}
\\
&=
\dfrac{g^2_0}{2}
\left(
\hat{d}^{(e)}_\lambda
-
\braket{\hat{d}^{(e)}_\lambda}_0
\right)^2
\quad.
\end{align}
In contrast to Ref.\cite{haugland2020}, $z_\lambda$ is required to be positive such that the CS transformation removes the nuclear dipole component from light-matter interaction and DSE terms. 

\setcounter{equation}{0}
\renewcommand{\theequation}{\thesubsection.\arabic{equation}}
\subsection{Connections between Original and Renormalized QED-CC}
\label{sec.renorm_qedcc_ansatz}
We show that $\hat{U}_z\ket{\Psi^\mathrm{qed}_\mathrm{cc}}$ approximates $\ket{\tilde{\Psi}^\mathrm{qed}_\mathrm{cc}}$. To this end, we start from the left-hand side of Eq.\eqref{eq.cs_qedcc_ansatz}
\begin{align}
\ket{\tilde{\Psi}^\mathrm{qed}_\mathrm{cc}}
&=
e^{\hat{Q}}
\hat{U}_z
\ket{\Phi_0,0_c}
=
\hat{U}_z
\hat{U}^\dagger_z
e^{\hat{Q}}
\hat{U}_z
\ket{\Phi_0,0_c}
\quad,
\end{align}
where we exploited $\hat{U}_z\hat{U}^\dagger_z=1$, and subsequently perform a BCH expansion 
\begin{align}
\hat{U}^\dagger_z
e^{\hat{Q}}
\hat{U}_z
&=
e^{-\hat{G}_z}
e^{\hat{Q}}
e^{\hat{G}_z}
=
e^{\hat{Q}}
+
[e^{\hat{Q}},\hat{G}_z]
+
\dots
\label{eq.cs_bch}
\end{align} 
with generator
\begin{align}
\hat{G}_z
&=
z_\lambda
(
\hat{b}^\dagger_\lambda
-
\hat{b}_\lambda
)
\quad.
\end{align}
We thus find
\begin{align}
\ket{\tilde{\Psi}^\mathrm{qed}_\mathrm{cc}}
&=
\hat{U}_z
(
e^{\hat{Q}}
+
[e^{\hat{Q}},\hat{G}_z]
+
\dots
)
\ket{\Phi_0,0_c}
\quad,
\vspace{0.2cm}
\\
&=
\hat{U}_z
\ket{\Psi^\mathrm{qed}_\mathrm{cc}}
+
\hat{U}_z
(
[e^{\hat{Q}},\hat{G}_z]
+
\dots
)
\ket{\Phi_0,0_c}
\,,
\end{align}
where we recover the original QED-CC ansatz\cite{haugland2020} as zeroth-order approximation of the BCH expansion Eq.\eqref{eq.cs_bch}. 

\setcounter{equation}{0}
\renewcommand{\theequation}{\thesubsection.\arabic{equation}}
\subsection{Renormalized QED-CCSD-1-SD1 and Bubble Contractions}
\label{sec.diagrams}
\begin{figure}[hbt!]
\includegraphics[scale=1.0]{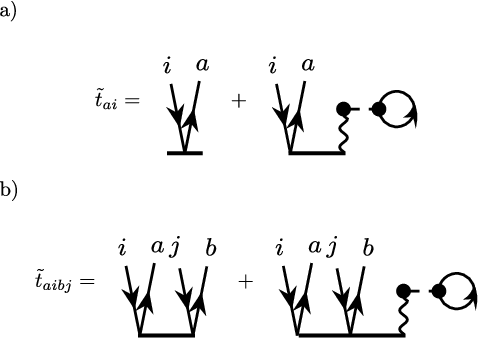}
\caption{Diagrams representing the CS-renormalized (a) singles amplitudes in Eq.\eqref{eq.renorm_singles} and (b) doubles amplitudes in Eq.\eqref{eq.renorm_doubles}.}
\label{fig.t1_t2_renorm}
\end{figure}

The polaritonic cluster operator for the QED-CCSD-1-SD1 scheme is obtained by augmenting Eq.\eqref{eq.qed_ccs_1_s1_cluster_op} with
\begin{align}
\hat{T}_2
=
\dfrac{1}{4}
t^{ab}_{ij}
\hat{E}_{ai}
\hat{E}_{bj}
\quad,
\quad
\hat{S}^1_2
=
\dfrac{1}{4}
s^\lambda_{aibj}
\hat{E}_{ai}
\hat{E}_{bj}
\hat{b}^\dagger_\lambda
\quad,
\end{align}
reflecting electronic doubles and mixed triples excitations with amplitudes, $t^{ab}_{ij}$ and $s^\lambda_{aibj}$, respectively. In CS-representation, the latter transforms as
\begin{align}
\hat{U}^\dagger_z
\hat{S}^1_2
\hat{U}_z
&=
\dfrac{1}{4}
\tilde{s}^\lambda_{aibj}
\hat{E}_{ai}
\hat{E}_{bj}
\left(
\hat{b}^\dagger_\lambda
+
\dfrac{
g_0
\braket{
\hat{d}_\lambda
}_0
}{
\sqrt{2\hbar\omega_c}
}
\right)
\quad,
\end{align}
which indicates a renormalization of electronic doubles amplitudes in analogy to Eq.\eqref{eq.renorm_singles} as
\begin{align}
\tilde{t}_{aibj}
&=
\tilde{t}^{ab}_{ij}
+
\dfrac{
g_0
\braket{
\hat{d}_\lambda
}_0
}{
\sqrt{2\hbar\omega_c}
}
\tilde{s}^\lambda_{aibj}
=
\tilde{t}^{ab}_{ij}
+
\tilde{s}^{ab}_{ij}
\quad,
\label{eq.renorm_doubles}
\end{align}
with an effective doubles contributions, $\tilde{s}^{ab}_{ij}$, respectively. In Fig.\eqref{fig.t1_t2_renorm}, we represent Eqs.\eqref{eq.renorm_singles} and \eqref{eq.renorm_doubles} in terms of the diagrammatic QED-CC notation discussed by Monzel and Stopkowicz\cite{monzel2024}.
Bubble contractions result here from the contribution of the polarization-projected mean-field dipole expectation value, which reads for effective electronic singles amplitudes (analogous contraction for $\tilde{s}^\lambda_{aibj}$ in Eq.\eqref{eq.renorm_doubles})
\begin{align}
\braket{\hat{d}_\lambda}
s^\lambda_{ai}
&=
\sum_k
\left(
d^{kk}_\lambda
+
\dfrac{\hat{d}^{(n)}_\lambda}{N_e}
\right)
s^\lambda_{ai}
\quad,
\end{align}
where our notation differs slightly from Ref.\cite{monzel2024}.
\begin{figure}[hbt!]
\includegraphics[scale=1.0]{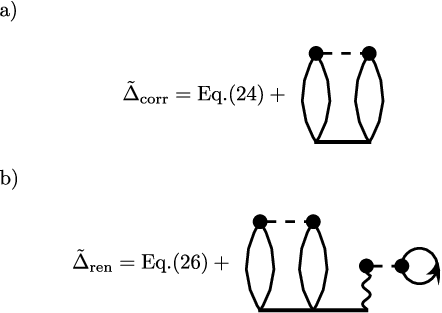}
\caption{Diagrams representing the a) connected doubles correction to the electronic correlation energy and b) the corresponding renormalization contribution given in Eq.\eqref{eq.renorm_doubles}.}
\label{fig.corr_renorm_t2_s2}
\end{figure}
Correlation and renormalization corrects acquire (effective) connected doubles corrections 
\begin{align}
\tilde{\Delta}_\mathrm{corr}
&=
\text{Eq.\eqref{eq.qed_cc_corr}}
+
\dfrac{1}{4}
\tilde{t}^{ab}_{ij}
\bar{\omega}^{ai}_{bj}
\quad,
\vspace{0.2cm}
\\
\tilde{\Delta}_\mathrm{ren}
&=
\text{Eq.\eqref{eq.qed_cc_renorm}}
+
\dfrac{1}{4}
\tilde{s}^{ab}_{ij}
\bar{\omega}^{ai}_{bj}
\quad.
\end{align}
as diagrammatically depicted in Fig.\eqref{fig.corr_renorm_t2_s2}.

\setcounter{equation}{0}
\renewcommand{\theequation}{\thesubsection.\arabic{equation}}
\subsection{Mean-Field Photon Number Expectation Value}
\label{sec.photonnr_qedhf}
The photon number operator in length-gauge representation is given by\cite{schaefer2020}
\begin{multline}
\hat{n}_\lambda
=
\dfrac{1}{\hbar\omega_c}
\left(
\hbar\omega_c
\hat{b}^\dagger_\lambda
\hat{b}_\lambda
-
g_0
\sqrt{\dfrac{\hbar\omega_c}{2}}
\hat{d}_\lambda
(
\hat{b}^\dagger_\lambda
+
\hat{b}_\lambda
)
\right)
\\
+
\dfrac{g^2_0}{2\hbar\omega_c}
\hat{d}^2_\lambda
-
\dfrac{1}{2}
\quad.
\label{eq.lg_photon_nr_op}
\end{multline}
The mean-field expectation value with respect to Eq.\eqref{eq.qedhf_ansatz} is given by
\begin{align}
\braket{\hat{n}_\lambda}_0
&=
\braket{
R
\vert
\hat{n}_\lambda
\vert
R}
=
\braket{
\Phi_0,
0_c
\vert
\hat{U}^\dagger_z
\hat{n}_\lambda
\hat{U}_z
\vert
\Phi_0,
0_c}
\quad,
\label{eq.lg_photon_nr_exp}
\end{align}
and expands after comparison with Eq.\eqref{eq.cs_polaritonic_hamilton} into
\begin{multline}
\braket{\hat{n}_\lambda}_0
=
\braket{
\Phi_0,
0_c
\vert
\hat{b}^\dagger_\lambda
\hat{b}_\lambda
\vert
\Phi_0,
0_c}
\\
-
\dfrac{g_0}{\sqrt{2\hbar\omega_c}}
\braket{
\Phi_0,
0_c
\vert
\Delta\hat{d}^{(e)}_\lambda
\left(
\hat{b}^\dagger_\lambda
+
\hat{b}_\lambda
\right)
\vert
\Phi_0,
0_c}
\\
+
\dfrac{g^2_0}{2\hbar\omega_c}
\braket{
\Phi_0,
0_c
\vert
(\Delta\hat{d}^{(e)}_\lambda)^2
\vert
\Phi_0,
0_c}
-
\dfrac{1}{2}
\quad,
\label{eq.lg_photon_nr_exp_explicit}
\end{multline}
where $\Delta\hat{d}^{(e)}_\lambda$ has been introduced in Eq.\eqref{eq.edip_corr}. The first and second line vanish identically and the third line is equivalent to Eq.\eqref{eq.cs_photon_expect}.
For systems with vanishing permanent dipole moment on the mean-field level of theory, the photon number expectation value is simply given by
\begin{align}
\braket{\hat{n}_\lambda}_0
&=
\dfrac{g^2_0}{2\hbar\omega_c}
\braket{
\hat{d}^2_\lambda
}_0
-
\dfrac{1}{2}
\quad,
\vspace{0.2cm}
\\
&=
\dfrac{g^2_0}{2\hbar\omega_c}
\left(
\braket{
\hat{d}^{(e)}_\lambda
\hat{d}^{(e)}_\lambda
}_0
+
2
\braket{
\hat{d}^{(e)}_\lambda
}_0
\hat{d}^{(n)}_\lambda
+
\hat{d}^{(n)}_\lambda
\hat{d}^{(n)}_\lambda
\right)
-
\dfrac{1}{2}
\,.
\nonumber
\end{align}
where both electronic and nuclear components of the dipole operator contribute.

\setcounter{equation}{0}
\renewcommand{\theequation}{\thesubsection.\arabic{equation}}
\subsection{Low-Frequency Limit of the QED-HF Reference State}
\label{sec.lowfreq_qedhf}
The QED-HF reference state in Eq.\eqref{eq.qedhf_ansatz} is a product state, $\ket{R}=\ket{\Phi_0}\otimes\hat{U}_z\ket{0_c}$, where the CS transformed cavity component can equivalently be written as
\begin{align}
\hat{U}_z
\ket{0_c}
&=
e^{-\frac{z^2_\lambda}{2}}
\sum_{n_c}
\dfrac{z^{n_c}_\lambda}{\sqrt{n_c!}}
\ket{n_c}
\quad,
\end{align}
with CS-parameter, $z_\lambda$, given in Eq.\eqref{eq.cs_transform}. In the low-frequency limit, we find $z_\lambda\to\infty$ as $\omega_c\to0$ and therefore
\begin{align}
e^{-\frac{z^2_\lambda}{2}}
\dfrac{z^{n_c}_\lambda}{\sqrt{n_c!}}
\to
0
\quad,
\quad
z_\lambda\to\infty
\quad,
\end{align}
independent of the photon number, $n_c$, such that $\hat{U}_z\ket{0_c}\to 0$ and $\ket{R}$ becomes undefined.



\end{document}